\documentclass[sigconf]{acmart}

\copyrightyear{2023}
\acmYear{2023}
\setcopyright{rightsretained}
\acmConference[HPDC '23]{Proceedings of the 32nd International Symposium on
High-Performance Parallel and Distributed Computing}{June 16--23,
2023}{Orlando, FL, USA}
\acmBooktitle{Proceedings of the 32nd International Symposium on
High-Performance Parallel and Distributed Computing (HPDC '23), June 16--23,
2023, Orlando, FL, USA}
\acmDOI{10.1145/3588195.3595947}
\acmISBN{979-8-4007-0155-9/23/06}

\usepackage{enumitem}
\usepackage{algorithm}
\usepackage{algpseudocode}
\usepackage{multicol}
\usepackage{subfigure}
\usepackage{caption}
\usepackage{booktabs} 
\usepackage{siunitx} 
\usepackage{etoolbox}

\usepackage[font=small,skip=0pt]{caption}

\usepackage{amsmath,algorithm,tabularx}
\usepackage{algpseudocode}

\makeatletter
\newcommand{\multiline}[1]{%
  \begin{tabularx}{\dimexpr\linewidth-\ALG@thistlm}[t]{@{}X@{}}
    #1
  \end{tabularx}
}
\makeatother
\AtBeginDocument{%
  \providecommand\BibTeX{{%
    \normalfont B\kern-0.5em{\scshape i\kern-0.25em b}\kern-0.8em\TeX}}}
\setcopyright{none}
\begin{document}

\renewcommand\footnotetextcopyrightpermission[1]{} 

\newcommand{\yujia}[1]{\textcolor{red}{#1}} 
\newcommand{\shixun}[1]{\textcolor{blue}{#1}} 
\title{FT-GEMM: A Fault Tolerant High Performance GEMM Implementation on x86 CPUs}

\author{Shixun Wu}
\email{swu264@ucr.edu}
\affiliation{%
  \institution{University of California, Riverside}
  \city{Riverside}
  \state{CA}
  \country{USA}
}
\author{Yujia Zhai}
\email{yzhai015@ucr.edu}
\affiliation{%
  \institution{University of California, Riverside}
  \city{Riverside}
  \state{CA}
  \country{USA}
}
\author{Jiajun Huang}
\email{jhuan380@ucr.edu}
\affiliation{%
  \institution{University of California, Riverside}
  \city{Riverside}
  \state{CA}
  \country{USA}
}
\author{Zizhe Jian}
\email{zjian106@ucr.edu}
\affiliation{%
  \institution{University of California, Riverside}
  \city{Riverside}
  \state{CA}
  \country{USA}
}
\author{Zizhong Chen}
\email{chen@cs.ucr.edu}
\affiliation{%
  \institution{University of California, Riverside}
  \city{Riverside}
  \state{CA}
  \country{USA}
}
\begin{abstract}

   General matrix/matrix multiplication (GEMM) is crucial for scientific computing and machine learning. However, the increased scale of the computing platforms raises concerns about hardware and software reliability. In this poster, we present FT-GEMM, a high-performance GEMM being capable of tolerating soft errors on-the-fly. We incorporate the fault tolerant functionality at algorithmic level by fusing the memory-intensive operations into the GEMM assembly kernels. We design a cache-friendly scheme for parallel FT-GEMM. Experimental results on Intel Cascade Lake demonstrate that FT-GEMM offers high reliability and performance -- faster than Intel MKL, OpenBLAS, and BLIS by 3.50\%$\sim$ 22.14\% for both serial and parallel GEMM, even under hundreds of errors injected per minute.

\end{abstract}

\settopmatter{printacmref=true}
\settopmatter{printfolios=true}

\maketitle
\section{Introduction}
 Due to performance-enhancing technologies, processor chips are more susceptible to transient faults. Transient faults can alter a signal transfer or corrupt the bits within stored values silently. As a consequence, reliability has been identified by the U.S. Department of Energy as one of the major challenges for exascale computing. We restrict our scope to fail-continue errors, from computing logic units (e.g., 1+1=3), namely soft errors. Several fault tolerance schemes have been proposed for the core computing routine GEMM to tolerate soft errors with low overhead. However, these methods are less efficient when using AVX-512-enabled processors given the huge gap between computation and memory transfer speed. In this poster, we present FT-GEMM, a high-performance GEMM being capable of tolerating soft errors on-the-fly \cite{zhai2021ft,wu2023ft,huang1984algorithm}. We compare our implementations with state-of-the-art GEMM implementations OpenBLAS \cite{openblasdtrsv}, BLIS, and Intel MKL on Intel Cascade Lake processors. Our main contributions include:

\begin{itemize}[leftmargin=*]
   \item GEMM using AVX-512 assembly instructions with a better performance (3.33\%-22.19\%) than the OpenBLAS, BLIS, and MKL.
   \item FT-GEMM with a negligible overhead (0.35\%-3.10\%) by fusing the memory-intensive operations, checksum encoding, and verification, into the GEMM assembly.
   \item  Cache-friendly multi-thread FT-GEMM with a negligible overhead (0.16\%-3.53\%).
   \item High performance (outperforms OpenBLAS, BLIS, and MKL by 3.5\%-22.1\%) and reliability under hundreds of errors injected. 
\end{itemize}

\begin{figure}[ht]
\vspace{-1mm}
\centering
\includegraphics[width=0.46\textwidth]{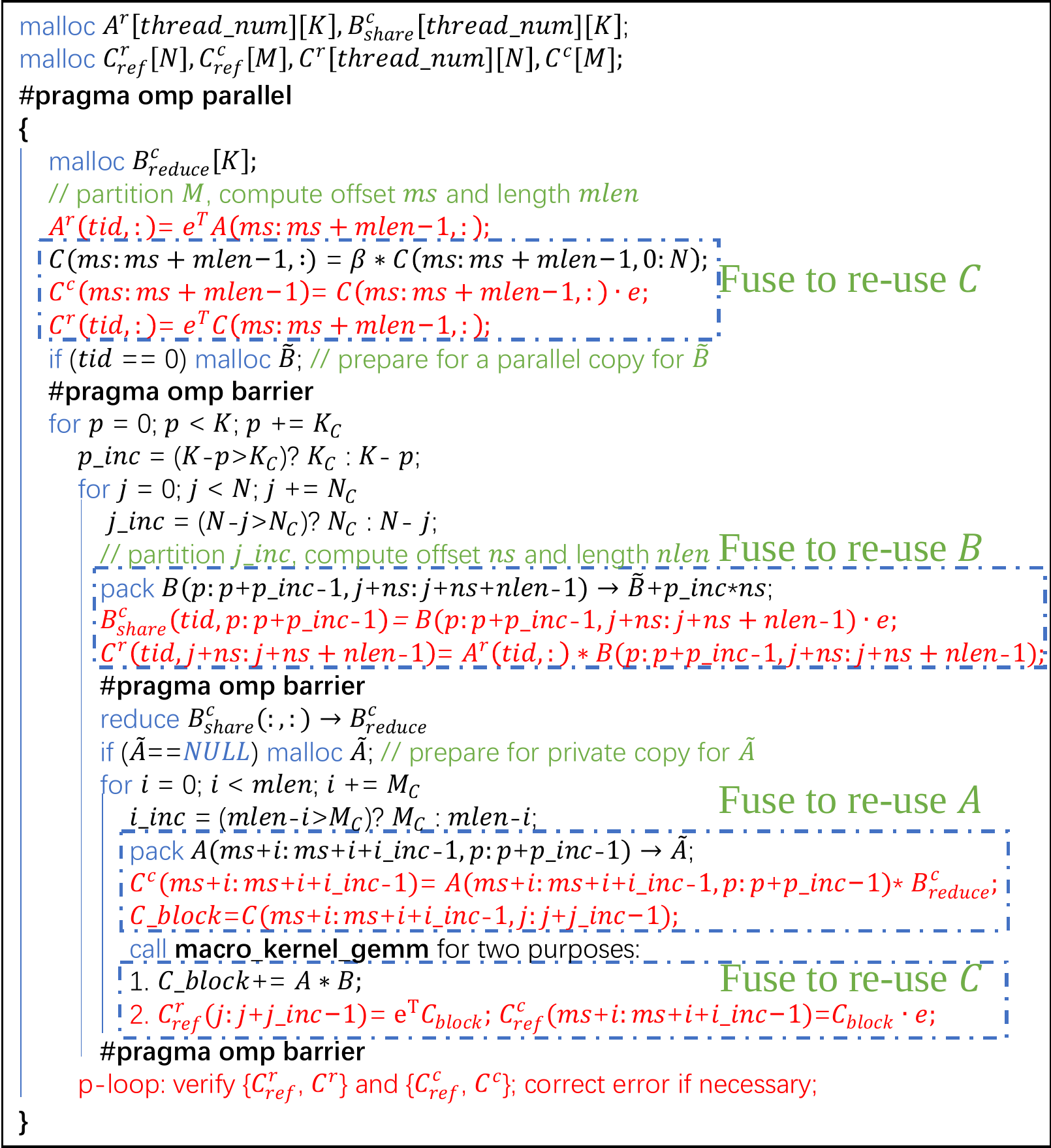}
\caption{Parallel FT-GEMM. ABFT \cite{zhai2021ft} operations are in red.}
\label{fig:dgemm-para-abft}
\end{figure}
\section{FT-GEMM}
\subsection{Implementation of DGEMM} We adopt packing and cache-blocking frames. The outermost three layers of the \verb|for| loop are partitioned to allow submatrices of $A$ and $B$ to reside in specific cache layers. The step sizes of these three \verb|for| loops, $M_C$, $N_C$, and $K_C$, define the shape of the macro kernel, which is determined by the size of each layer of the cache. A macro kernel updates an $M_C\times N_C$ submatrix of $C$ by iterating over $A$ $(M_R\times K_C)$ multiplying $B$ $(K_C\times N_R)$ in micro kernels.

\subsection{FT-DGEMM}
As discussed in the previous section, the huge gap between memory transfer and floating-point computation is the reason the $O(n^2)$ checksum-related operations can no longer be amortized by $O(n^3)$ GEMM. We therefore design a fused ABFT \cite{zhai2021ft} scheme to minimize the memory footprint of checksum operations. To be more specific, the encoding of $C^c$ and $C^r$ is fused with the matrix scaling routine $C$=$\beta C$. When we load $B$ to pack it to the continuous memory buffer $\Tilde{B}$, checksum $B^c$ and checksum $C^r$ are computed simultaneously by reusing $B$. In this fused packing routine, each $B$ element is reused three times for each load. Similarly, each element of $A$ loaded for packing is reused to update the column checksum $C^c$. In the macro kernel, which computes $C_{block}$+=$\Tilde{A}\cdot\Tilde{B}$, we reuse the computed $C$ elements at register level to update the reference checksums $C^r_{ref}$ and $C^c_{ref}$ in order to verify the correctness of the computation. By fusing the ABFT memory footprint, the FT overhead becomes purely computational, decreasing from about 15\% to 2.94\%.

\subsection{Parallel FT-DGEMM}
In addition to providing highly efficient serial implementations, we further enable the multithreading support for DGEMM with and without fault tolerance. On Intel Cascade Lake server CPUs, physical cores share a large unified L3 cache while each physical core holds a smaller private L2 cache. To map this cache hierarchy in a threaded implementation, we allocate a memory buffer shared among all the threads for $\Tilde{B}$, and each thread requests a private memory buffer for $\Tilde{A}$. The computation workload on the $C$ matrix is partitioned along the $M$-dimension. Since memory buffers $\Tilde{A}$ are thread-private, each thread packs data from matrix $A$ into their own $\Tilde{A}$ buffers. When packing matrix $C$ into the shared memory buffer $\Tilde{B}$, the memory access workloads are partitioned along the $N$-dimension and each thread is responsible for packing a chunk of $\Tilde{B}$. We conduct checksum encoding for the row checksum vector of $A$ ($A^r$) and full checksum vectors of $C$ ($C^c, C^r$). To compute the $C$ checksums, we partition the $C$ matrix along the $M$-dimension such that each thread computes a slice of the column checksum $C^c$ while maintaining a local copy of its own row checksum vector $C^r$. Similarly, we partition the $A$ matrix along the $M$-dimension to compute its row checksums $A^r$ in parallel. The checksum encoding of $B^c$ is fused with the parallel packing operation for $B$ to $\Tilde{B}$ and simultaneously, we update the reference row checksum of $C$. Therefore, each $B$ element loaded from the main memory is re-used three times. Since the parallel copy operation partitions $B$ from the $N$-dimension, an extra stage of reduction operation among threads is required to compute the final column checksum $B^c$. 

\section{Experimental Evaluation}
\label{section:eva}

To validate the effectiveness of our optimizations, we compare the performance of FT-BLAS with three state-of-the-art BLAS libraries: Intel oneMKL (\verb|2020.2|, abbreviated as MKL in this Section), OpenBLAS (\verb|0.3.13|), and BLIS (\verb|0.8.0|), on an Intel Xeon W-2255 Cascade processor equipped with 3.70 GHz base frequency and 32 GB DDR4-2933 RAM. Hardware prefetchers is enabled according to the Intel BIOS default. We repeat each measurement twenty times and then report the average performance. The performance is averaged for matrices ranging from $2048^2$ to $10240^2$. For the multi-threading parallel benchmark, we test the matrices ranging from $512^2$ to $20480^2$. We compile the code with \verb|icc 19.0| and the optimization flag \verb|-O3|.

\setlength{\subfigcapskip}{-5pt}
\setlength{\subfigbottomskip}{0pt}
\begin{figure}[t] 
\centering
\subfigure[FT-DGEMM, Serial]
{
\includegraphics[width=0.21\textwidth]{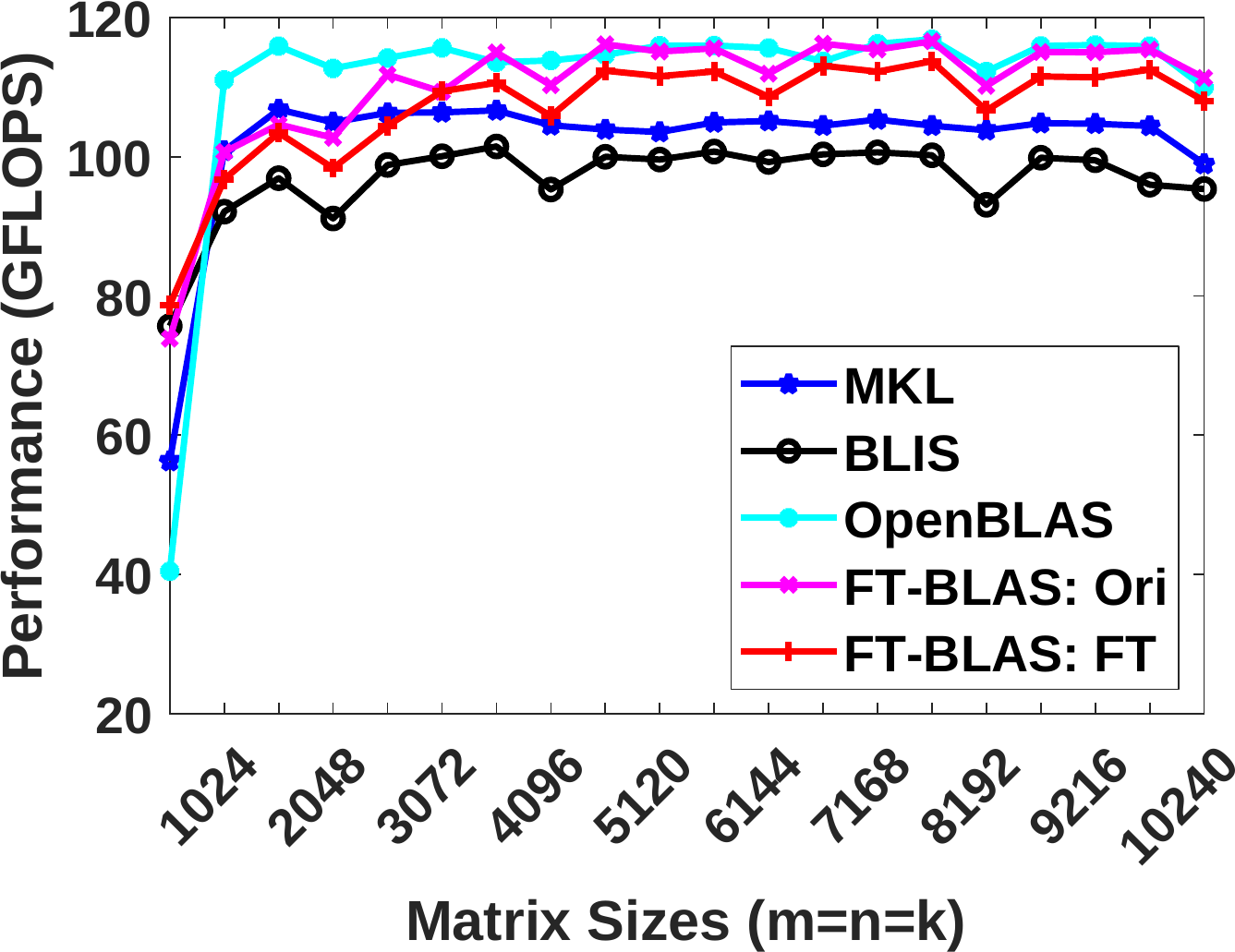}

}
\subfigure[FT-DGEMM,Paralllel]
{
\includegraphics[width=0.21\textwidth]{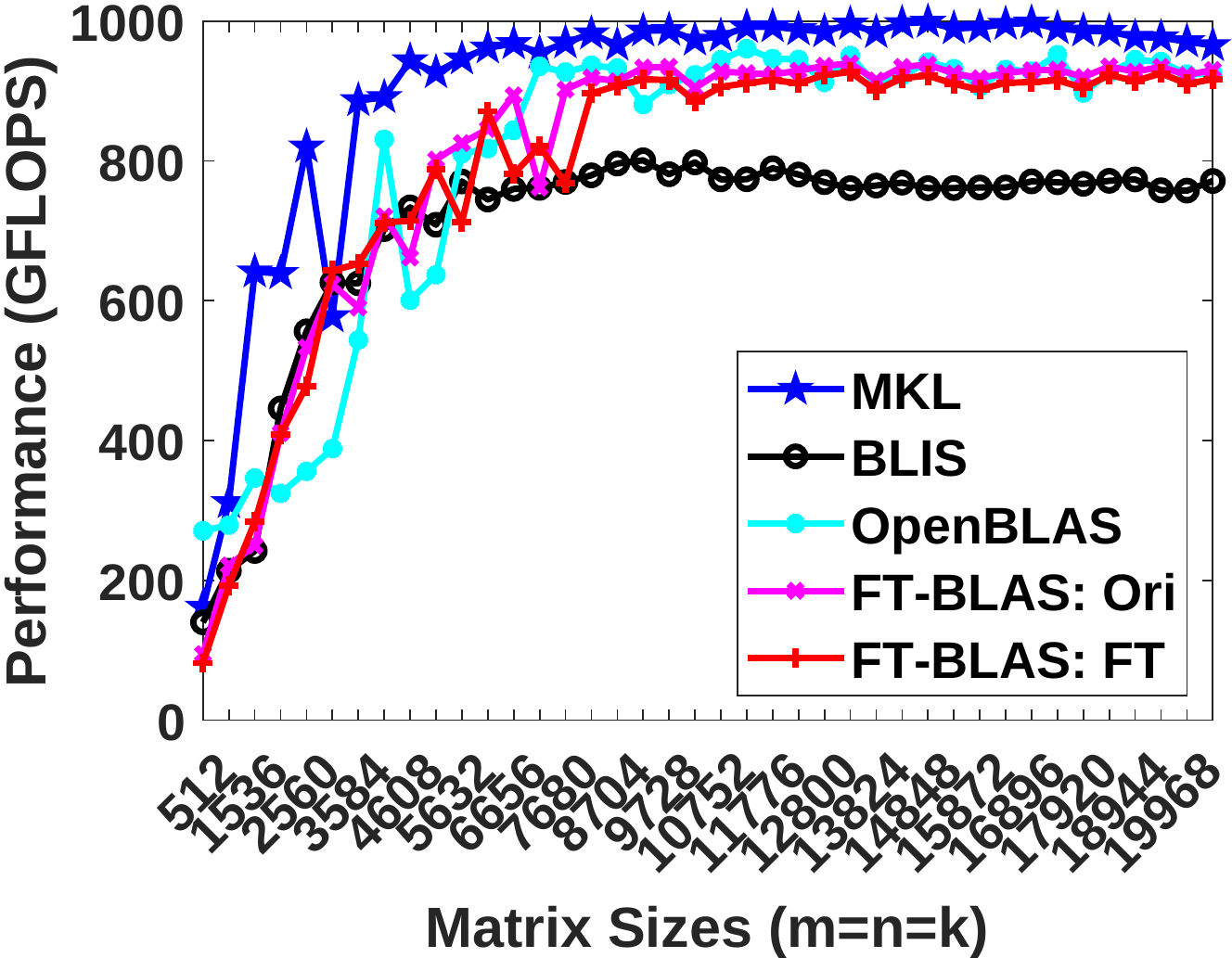}
}

\subfigure[Error injection, Serial]
{
\includegraphics[width=0.21\textwidth]{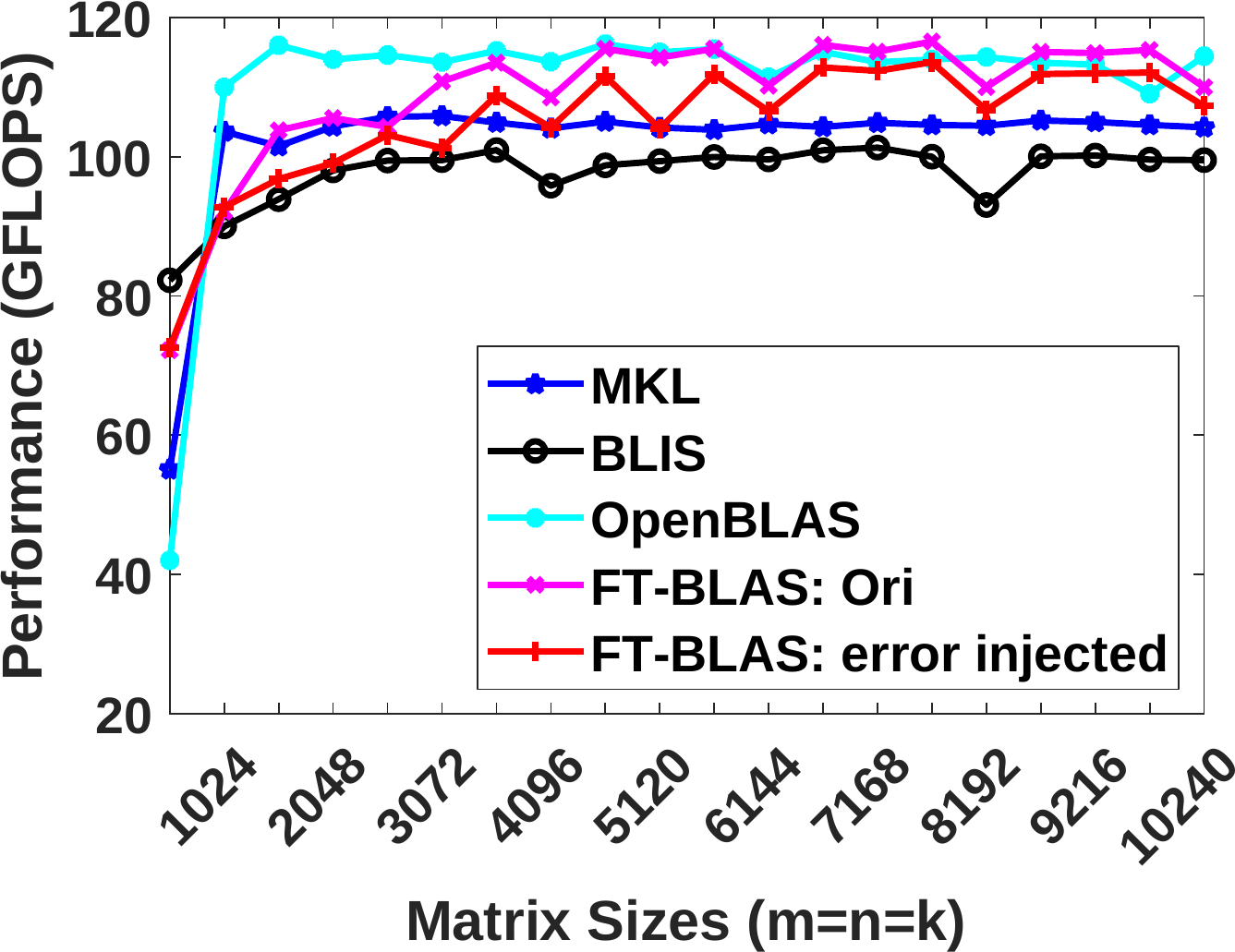}
}
\subfigure[Error injection, Parallel ]
{
\includegraphics[width=0.21\textwidth]{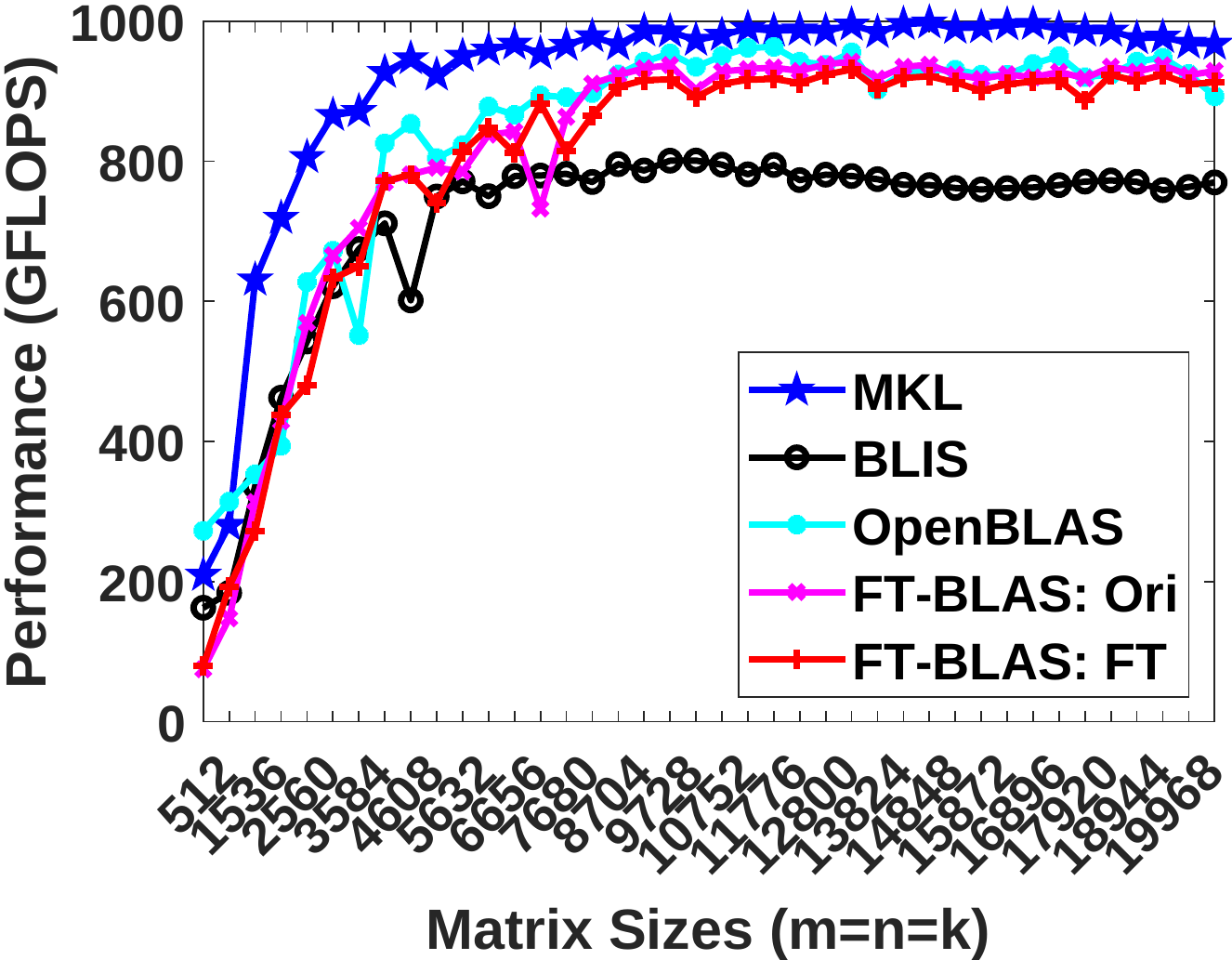}
}
\caption{Comparisons of FT-GEMM on Intel Cascade Lake.}
\label{fig:ft-blas-para}
\end{figure}
\subsection{Performance of FT-GEMM}
The results in this section were obtained with fault tolerant DMR and ABFT operating, but not under active fault injection. In Fig. \ref{fig:ft-blas-para}(a), our baseline GEMM implementations (FT-GEMM: Ori) present comparable or better performance compared with MKL, OpenBLAS, and BLIS. Our fused fault tolerant strategy for compute-bound DGEMM generates 1.17\%-3.58\% overhead on average over the baseline. Fig. \ref{fig:ft-blas-para}(b) compares the parallel performance of FT-GEMM with FT capability. With the scalable parallel design and ABFT operations fused into packing routines and assembly kernels, FT-DGEMM presents a negligible overhead (1.79\%). The performance of our DGEMM with FT is 16.97\% faster than BLIS, comparable to OpenBLAS while slightly underperforming the close-sourced Intel MKL.
\subsection{Performance under error injection}
We validate the effectiveness of our fault-tolerance scheme by injecting multiple computing errors into each of our computing kernels and verifying our final computation results against MKL. External error injection tools often significantly slow down the native program. Therefore, we inject errors at the source code level to minimize the performance impact on native programs. In Fig. \ref{fig:ft-blas-para}(c), our protection scheme surpasses OpenBLAS and BLIS by 22.89\% and 21.56\% and the closed-source MKL by 4.98\% even while tolerating 20 injected errors. In Fig. \ref{fig:ft-blas-para}(d), our FT-BLAS presents a performance comparable to OpenBLAS and is 16.83\% faster than BLIS.

\section{Acknowledgement}
This work was supported by the U.S. Department of Energy, Office of Science, Office of Advanced Scientific Computing Research, Scientific Discovery through the Advanced Computing (SciDAC) program under Award Number DE-SC0022209.

\bibliographystyle{ACM-Reference-Format}
\balance
\bibliography{bib/refs}

\end{document}